\documentclass[aps,floatfix,twocolumn,amsmath]{revtex4}
\usepackage{graphicx}
\newcommand{\be}{\begin{equation}}
\newcommand{\ee}{\end{equation}}
\newcommand{\bea}{\begin{eqnarray}}
\newcommand{\eea}{\end{eqnarray}}
\newcommand{\no}{\noindent}

\begin{document}


\title{Vibrating Winding Branes, Wrapping Democracy \\and\\Stabilization of Extra Dimensions \\ in \\ Dilaton Gravity}


\author{Tongu\c{c} Rador}
\email[]{tonguc.rador@boun.edu.tr}

\affiliation{Bo\~{g}azi\c{c}i University Department of Physics \\ 34342 Bebek, \.{I}stanbul, Turkey}


\date{\today}

\begin{abstract}

We show that, in the context of dilaton gravity, a recently proposed  
democratic principle for intersection possibilities of branes winding around extra dimensions yield stabilization, even with the inclusion of momentum modes of the wrapped branes on top of the winding modes. The constraints for stabilization massaged by string theory inputs 
forces the number of observed dimensions to be three. We also discuss consequences of adding ordinary matter living in the observed dimensions.

\end{abstract}


\maketitle

\section{Introduction}

In this work we outline a dynamical scenario for the stabilization of extra dimensions which are required for the consistency of string theory. Stabilization  of extra dimensions is rather crucial because we know that the observed universe is very large. Therefor it is a puzzle that other dimensions remained comparatively small in contrast to the fact that there are many indications that the observed universe was expanding in its entire history, and in the inflationary period very fast too. 

In \cite{art1} a simple mechanism was proposed to achieve this result and in \cite{art2}-\cite{art4} some extensions were considered. The mechanism proposed was that the winding modes of a brane wrapped around a compact dimension provide a resisting force to expansion much like rubber band
would if one wraps it around balloon and inflate. This is very close the idea once proposed by Brandenberger and Vafa \cite{art5} , and also fits in the context of "Brane Gas Cosmology" \cite{art6}-\cite{art23}. 

In \cite{art4}, where stabilization was shown to occur within the context of dilaton gravity, the brane wrapping configuration was such that there was a gas of branes wrapping around the entire set of extra dimensions which was a six dimensional tori. This work is aiming at the study of what might happen for arbitrary decimations (partitionings) of the extra dimensions to product manifolds each of which is again assumed to be tori. However for an arbitraty decimation this would be a very difficult task. 

Recently to study more general winding configurations and hence generalize the mass of intuition gained we introduced \cite{art5a} the idea of democratic wrapping in which all the possible brane windings are allowed. Here we generalize the results of \cite{art5a} to dilaton gravity. After the introduction of the main formalism we study a system where the decimation of the extra dimensions will be done symmetrically, that is the number of extra dimensions is
taken to be $Np$ where $N$ is the order of decimation and $p$ is the dimension of one unit of the partitioning. We first consider winding modes only and show that stabilization of extra dimensions is achieved. The result does not depend on $p$ and the uniqueness of the solutions hints at a general solution 
when the decimation is not symmetric. We then proceed to the inclusion of  momentum (vibration) modes of the branes. Here the condition for stabilization initially depends on $p$ but further considerations (such as requiring the observed dimensions to be expanding) makes this dependence evaporate and again hints at a general solution.

We also show that adding presureless matter living in the observed space does not alter the results. We however omit the general study for radiation and vacuum energy dominated eras of the universe.

\section{Formalism}

For cosmological purposes we can take the metric to be,
\be
ds^{2}=-dt^{2}+e^{2 B(t)} dx^{2} + \sum_{i} e^{2 C_{i}(t)} dy_{i}^{2}\;.
\ee
\no Here the $C_{i}$ and $y_{i}$ represent the scale factor and the coordinates of extra dimensions respectively. The dimensionality of each partition is $p_{i}$. For clarity we separated the observed dimensions with scale factor $B$ and dimensionality $m$. We also assume that the observed dimensions $x_{i}$ are non-compact  and the $y_{i}$ represent compact extra dimensions with the topology of tori each. The total space time dimensionality is $d=1+m+\sum_{i}p_{i}$.

We take the action in the presence of dilaton coupled to matter to be,

\be
S=\frac{1}{\kappa^{2}}\int dx^{d}\;\sqrt{-g}\;e^{-2\phi}\left[R + 4 (\nabla \phi)^2 + e^{a\phi}\mathcal{L}_{m}\right]\;.
\ee

In \cite{art4} it has been shown \footnote{In \cite{art4} the fluxes $H_{\mu\nu\sigma}=0$. We follow the same approach here}, in a cosmological context where the only variable can be taken to be time and with the assumption that the matter lagrangian $\mathcal{L}_{m}$ follows a
hydrodynamical dust form, that the equation of motions are given by

\begin{subequations}
\bea
R_{\mu\nu}+2\nabla_{\mu}\nabla_{\nu}\phi &=& e^{a\phi}\left[T_{\mu\nu}-(\frac{a-2}{2})\rho g_{\mu\nu}\right] \\
R+4 \nabla^{2}\phi -4 (\nabla\phi)^{2} &=& -(a-2)e^{a\phi}\rho\;.
\eea
\end{subequations}

\no Here $\rho$ is the sum of energy densities of various contributions and for hydrodynamical dust, the energy momentum tensor for each contribution will have the usual "pressure equals a coefficient times the corresponding energy density" form. The fact that we can choose a continuous energy density for the contributions of branes to the energy-momentum tensor is because we take a gas of them spread over the observed dimensions $m$, (For further explanations, see \cite{art1}-\cite{art3}). As evident from notation $T_{\mu\nu}$ is the sum of all contributions. Also for the sake of clarity we must emphasize that each contribution to $T_{\mu\nu}$  are
separately conserved.

With these assumptions the equations of motion for the scale factors can be cast as follows (we set $\kappa^{2}=1$),

\begin{subequations}
\bea
\ddot{B}&=&-k\dot{B}+e^{a\phi}\left[T_{\hat{b}\hat{b}}-\tau\rho\right]\;,\\
\ddot{C_{i}}+k\dot{C_{i}}&=&e^{a\phi}\left[T_{\hat{c}_{i}\hat{c}_{i}}-\tau\rho\right]\;,\label{cieq}\\
\ddot{\phi}&=&-k\dot{\phi}+\frac{1}{2}e^{a\phi}\left[T-(d-2)\tau\rho\right]\;,\\
k^{2}&=&m\dot{B}^{2}+\sum_{i}p_{i}\dot{C_{i}}^{2} + 2 e^{a\phi}\rho\;,\label{aninki} \\
k&\equiv& m\dot{B}+\sum_{i}p_{i}\dot{C_{i}}-2\dot{\phi}\;.
\eea
\end{subequations}

\no The hatted indices refer the the orthonormal coordinates and $T$ represents the trace of the total energy-momentum tensor. Here we have also defined $\tau=(a-2)/2$ for a compact notation.

The energy momentum tensor for each contribution can be found using the conservation equation $\nabla^{\mu}T^{\alpha}_{\mu\nu}=0$ (here $\alpha$ is an index for describing the separate contributions to the total energy-momentum tensor $T_{\mu\nu}$) and yields the known result,

\be
\rho^{\alpha}=\rho_{0}^{\alpha} \exp\left[-m B +\sum_{i}(1+\omega_{i})p_{i}C_{i})\right]\;.
\ee

\no For brane winding modes $w_{i}=-1$ and $w_{i}=0$  along winding and non-winding (transverse) directions respectively \footnote{This has been shown to be the case in \cite{art1} for a gas of branes, and used exclusively in \cite{art2}-\cite{art4}}.

If there is stabilization of the extra dimensions, all $C_{i}$ have to be constants and the remaining equations for $B$ and $\phi$ will have to be satisfied. The impact of stabilization 
on these equations can be summarized as

\be
\sum_{i} p_{i} \left[T_{\hat{c}_{i}\hat{c}_{i}}-\tau\rho\right]=0,
\ee

\no which yields 

\be
T=-\left[1-(d-1-m)\tau\right]\rho\;.
\ee

Finally, since the observed space is transverse to all the winding modes, we have $T_{\hat{b}\hat{b}}=0$. This will yield the following set of equations, which could be dubbed as the stabilization constrained equations

\begin{subequations}
\label{conseq}
\bea
\ddot{B}+k\dot{B}&=&-\tau\sigma e^{a\phi-mB}\;,\\
\ddot{\phi}+k\dot{\phi}&=&-\frac{1}{2}\sigma\left[1+(m-1)\tau\right] e^{a\phi-mB}\;,\\
k^{2}&=&m\dot{B}^{2}+2\sigma e^{a\phi-mB}\;.
\eea 
\end{subequations}

\no Here
 
\be\label{defsig}
\sigma=e^{mB(0)} \sum_{\alpha} \rho^{\alpha}(0)\;,
\ee

\no is a constant that depends on the stabilized values of the $C_{i}$ and the density parameters $\rho_{0}^{\alpha}$. We obviously must have $\sigma>0$.

\no We postpone the study of these equations until we establish further conditions for stabilization.

\section{N-Fold Symmetric Decimation}

In \cite{art5a} it was proposed that if there exists solutions for stabilization in a symmetric
decimation of extra coordinates in the form of $d-1-m=N p$, then this hints at a solution for
a general decimation provided the conditions do not depend on $p$. The advantage of this 
approach is that all the $C_{i}$ equations effectively reduce to a single equation and this 
facilitates the study of stabilization to a considerable extent. 

We will also consider what was called democratic winding scheme in \cite{art5a}. Namely wrappings of all
possible kinds are allowed. This can be better exposed as follows

\bea
(p)pppp... &\oplus& \;\; {\rm N sources}\oplus \nonumber\\
(pp)ppp... &\oplus& \;\; {\rm N sources}\oplus \nonumber\\
(ppp)pp... &\oplus& \;\; {\rm N sources}\oplus \nonumber\\
{\rm all \;\; N-1\;\;modes}&\vdots& \;\;{\rm N sources}\oplus \nonumber\\
 (ppp....) && \;\; {\rm 1 source}\nonumber
\eea

\no where a parenthesis means there is a brane wrapping over that particular subelement of the decimation. For example what we mean by the first line in the above is that there are branes around $(p)pp...$, $p(p)p...$, $pp(p)...$ and so on until we have the last element of the decimation is also wrapped. There would be $N$ such branes and hence $N$ energy-momentum tensors with equal coefficients, e.g. all of
them will have $\rho_{0}^{p}$. Similarly for the windings around two subelements we have all the
combinations of the form $(pp)pp...$, $p(pp)p...$, $pp(pp)...$ and so on, and all the coefficients of the energy momentum tensors corresponding to those wrappings will be $\rho_{0}^{pp}$. Democracy mandates we should allow all such combinations until we end up with a single brane wrapping around the whole $Np$ dimensional space.

With these assumptions the condition for stabilization is simplified to the following

\begin{subequations}
\bea
&-&\sum_{n=0}^{N-1} \alpha_{n} \xi_{n}  X^{n}=0 \label{poly1}\;, \\
\xi_{n}&=&\left[N(1+\tau)-n\right]\label{poly1a}\;,
\eea
\end{subequations}

\no with $X=e^{-pC}$. Here $\alpha_{n}$'s are directly related to the energy coefficients. For example $\alpha_{N-1}=\rho_{0}^{p}$, $\alpha_{N-2}=\rho_{0}^{pp}$ and so on until $\alpha_{2}$. The coefficient $\alpha_{1}$ on the other hand is equal to the coefficient of the energy-momentum tensor which wraps around the whole $Np$ normalized by $N$, this is because we need only one such term and we wanted to have a compact form for (\ref{poly1}).

The resolution of stabilization is to find solutions of (\ref{poly1}), without any contrived choices of $\alpha_{n}$'s, such that $X>0$. All we have to do is to find the number of  sign changes $s$ of the coefficients in (\ref{poly1a}) counting from the highest power term. 
Then Descartes' sign rule tells us that the number of positive roots is either $s$ or
smaller than $s$ by a multiple of $2$. Now, it is clear that there can be no such changes if $(1+\tau)<0$. Also since $n<N$ there can be no changes if $(1+\tau)>1$. Therefor  to have a sign change we must at least have,

\be
-1<\tau<0 \Longrightarrow 0<a<2\;.
\ee

\no this is the sine-qua-non for stabilization but it is not sufficient. To ensure a sign change we should also impose the weakest condition on the coefficients (\ref{poly1a}). The one that will first change sign is the coefficient of the largest power $\xi_{N-1}$. Demanding it to be negative will yield

\be
N\tau <-1\;.
\ee

\no These conditions together limit the possibilities for $\tau$ and the number of partitions $N$ in a decimation, which we could also call the decimation order. It is interesting that these seemingly unrelated quantities
are tied in this way. For example, if we take our branes to be $D$-branes, then string theory  fixes $\tau=-1/2$ ($a=1$) which would mean $N>2$. However as we will see shortly this condition will not be required if we consider brane momentum modes as well. Finally, the monotonic decrease of the $\xi_{n}$ by $n$ means that there can be only one sign change and hence only one solution. Since the result does not depend on $p$ we may hope for a general solution where the decimation is done asymmetrically. 

At this point we would like to remark that in \cite{art5a}, where the discussion was confined to Einstein gravity, the stabilization conditions were on the number of observed dimensions $m$. Here the condition on $m$ will reappear as we will see in section \ref{seceq} when we consider the stabilization constrained equations of motion (\ref{conseq}). 

\subsection{Adding Momentum Modes}

Having established the conditions for stabilization with the winding modes we can also look for the impact  of the momentum modes. The democratic winding scheme leads to a very interesting simplification here. Let us expose two cases explicitly to understand the behavior better. 

First, consider the mode where only one part of the decimation is wrapped by a brane. The pressure coefficient for such a mode is $1/p$ along the winding direction  and zero for the others. Therefor the energy density will have the form (omitting $e^{-mB}$ which would factorize)

\bea
e^{-p(1+\frac{1}{p})C}\times &\underbrace{e^{-pC}\cdots e^{-pC}}& =e^{-NpC}e^{-C}\;, \nonumber\\
&{\rm N-1\;terms}&\nonumber
\eea

\no and the factor of this contribution to the RHS of the $C_{i}$ 
equations of motion (\ref{cieq}) will be

\be
\tilde{\rho}_{o}^{p} \left[1\times \frac{1}{p} - N \tau\right] \;\;,
\ee

\no since there will be $N$ such energy densities along the other parts of the decimation. We have also introduced the convention that the $\tilde{\rho}_{o}$'s  refer to the momentum mode energy coefficients, as opposed to the fact that ones without tildes refer to the winding modes.

Second, consider the mode where only two parts of the decimation is wrapped by a brane. The pressure coefficient for such a mode is $1/2p$ along the winding directions and zero for the others. Therfore the energy density will have the form (omitting $e^{-mB}$ which would factorize)

\bea
e^{-p(1+\frac{1}{2p})C}\times e^{-p(1+\frac{1}{2p})C} &\underbrace{e^{-pC}\cdots e^{-pC}}& =e^{-NpC}e^{-C}\;, \nonumber\\
&{\rm N-2\;terms}&\nonumber
\eea

\no and the factor of this contribution to the RHS of the $C_{i}$ equations of motion (\ref{cieq}) will be

\be
\tilde{\rho}_{o}^{pp}\left[ 2\times \frac{1}{2p} - N \tau \right] = \tilde{\rho}_{o}^{pp} \left[ \frac{1}{p} - N \tau \right]\;.
\ee

This feature actually carries to all the other windings and we get the following for the stabilization 
condition when the momentum modes are added on top of the winding modes

\begin{subequations}
\bea
-\tilde{\alpha} \left[N\tau-\frac{1}{p}\right] X^{\frac{Np+1}{p}} - \sum_{n=0}^{N-1} \alpha_{n} \xi_{n}  X^{n}=0\label{poly2}\;,\\
\tilde{\alpha} = \tilde{\rho}_{o}^{p}+\tilde{\rho}_{o}^{pp}+\cdots+\tilde{\rho}_{o}^{pp\cdots p}/N\;.\nonumber
\eea
\end{subequations}

Since there is only one term with a noninteger power of $X$ in (\ref{poly2}), we can still consider it as a polynomial with integer powers of  $X^{p}$. It is clear that there can be no sign changes in (\ref{poly2}) if $N\tau>1/p$ and if $\tau<-1$. Therefor the constraint $\tau$ is 

\be
-1<\tau<\frac{1}{pN}\;\;.
\ee

\no Unlike the pure winding case the condition $\tau<1/(pN)$ guarantees that there is at least one sign change. As before however the condition $\tau>-1$ guarantees we remain in this regime.

\no{\bf{Adding ordinary matter:}} There is one further possibility we can easily consider: the presureless dust in the observed dimensions. This follows from the fact that since the pressure coefficients of this term are all vanishing the contribution
will be proportional to $e^{-mB}e^{-NpC}$. Adding this term to the RHS of the $C_{i}$ equations (\ref{cieq}) we get

\be\label{poly3}
-\tilde{\alpha} \left[N\tau-\frac{1}{p}\right] X^{\frac{Np+1}{p}} -\bar{\alpha} \tau X^{N}- \sum_{n=0}^{N-1} \alpha_{n} \xi_{n}  X^{n}=0\;,
\ee

\no where $\bar{\alpha}=\rho_{o}^{matter}$. It can be shown that the constraints on $\tau$ are unaltered with this modification. Also since $T_{\hat{b}\hat{b}}=0$ for ordinary presureless dust, the stabilization constrained equations of motion (\ref{conseq}) are also unaltered. Other forms of matter living in the observed dimensions will complicate things beyond the
scope of this work and are omitted.

It is intriguing that the upper bound on $\tau$ is like $\tau<\omega/N$ where $\omega$ is the pressure coefficient of the energy-momentum contributions that yield he highest power of $X$ in the stabilization equation while the lower bound follows that of the lowest order term, giving the largest allowed region for $\tau$. If we did not adopt the democratic scheme the constraints on $\tau$ would have been tighter.

\section{Equations for $B$ and $\phi$}\label{seceq}

Having established the conditions for the stabilization of the extra dimensions we can turn back to the equations (\ref{conseq}) that controls the evolution of $B$ and $\phi$. The equations could have many solutions  but it is possible to find a power law solution with the following ansatz \footnote{These many solutions to (\ref{conseq}) would give various transient behaviour for the early time evolution however the late time behaviour can always be sought in terms of a power-law ansatz since our analysis for stability does not depend on the initial conditions}

\begin{subequations}
\bea
B(t)\sim \beta \ln{t}\;,\\
\phi(t)\sim \varphi \ln(t)\;.
\eea
\end{subequations}

\no There is a solution to \ref{conseq} with the following

\begin{subequations}
\label{bphi}
\bea
\beta&=&-\frac{2\tau}{1+(m-1)\tau^{2}}\label{bphib}\;,\\
\varphi&=&\frac{-2+m\beta}{2(1+\tau)}=-\frac{1+\tau(m-1)}{1+(m-1)\tau^{2}}\;,\\
\sigma&=&2\frac{1-2\tau-(m-1)\tau^{2}}{(1+(m-1)\tau^{2})^{2}}\;.
\eea
\end{subequations}

There are various implications of these. First if we want the observed dimensions to expand we must have $\tau<0$ \footnote{The contracting case for the solution in (\ref{bphib}) 
might be interesting on its own for other purposes but we
do not pursue this in this paper.}. Also we do not want the dilaton to grow since we would like to use
this solution for late time cosmology and a growing dilaton would take us in the strong coupling regime. Finally $\sigma$ has to be positive following its definition (\ref{defsig}). The corresponding conditions on $\tau$ are
given in the same order as below 

\begin{subequations}
\bea
\tau<0\;,\\
-\frac{1}{m-1}\leq \tau\;.\\
\frac{1}{1-\sqrt{m}}<&\tau&<\frac{1}{1+\sqrt{m}}\;,
\eea
\end{subequations}

\no Combining everything and therfore ignoring redundant constraints our final result is

\be\label{finconstau}
-\frac{1}{m-1}\leq \tau<0\;.
\ee

We see that the condition that the observed space expands evaporated the $p$ dependence of the constraints on $\tau$, in view of this and the fact that there is only one positive solution to the stabilization equation we can hope for a solution in the general case of asymmetric decimation. 

If we insist on the fact that the objects we are dealing with are $D$-branes, string theory gives an unambiguous answer $\tau=-1/2$, (or $a=1$). This, in turn, puts a limit on the dimensionality of 
observed dimensions $m\leq3$. When $m=3$ the scale factor of the observed space evolves as $t^{2/3}$, exactly like presureless dust solution of standard cosmology. Furthermore with $m=3$ the dilaton is stabilized and we have the equivalence of the Einstein frame and the string frame. This result is not unexpected since as shown in \cite{art5a} within the context
of pure Einstein gravity $m=3$ is a sufficient (but in that context not necessary) condition for stabilization of extra dimensions using symmetric decimation and democratic wrapping \footnote{In \cite{art5a} the momentum modes were not considered. However as it is clear from this work the inclusion of the momentum modes, unexpectedly, does not alter things drastically.}.

\section{Stability of the equilibirum point}

Having established the equilibrium conditions and solutions, we should make sure that the equilibrium is
indeed stable. In this section we discuss this issue and show that all admissible (see discussion below) initial data will converge to the stabilization point. Much of this discussion follows in a very close way to the arguments presented \cite{art5a}. The equations for the scale factors of the extra dimensions is (within the symmetric decimation where there is only one such equation),

\be\label{equic}
\ddot{C}=-k\dot{C}+e^{a\phi-mB} P_{s}(X=e^{-pC})
\ee 
Here $P_{s}(X)$ is given by (\ref{poly3}). If we expand $C$ around the equilibrium point such that we keep only the linear perturbations $C=C_{0}+\delta{C}$ we would get the following

\be\label{equib}
\delta\ddot{C}=-k\delta\dot{C}-p\delta C e^{a\phi-mB}\;X_{0}\; P_{s}'(X_{0}),
\ee
where $P_{s}'(X_{0})$ denotes derivative of $P_{s}$ as the equilibrium point.

This is like a motion under two forces. The force which is proportional to $k$ can either be a driving
or a friction force depending on the sign of $k$. But as clear from the equations of motions, ${\rm Sign}(k)$  is a constant of motion  since $k$ is never allowed to vanish by  (\ref{aninki}). In \cite{art4} it has been observed that when ${\rm Sign}(k)<0$ one has a driving force and a singularity is reached in finite proper time. For ${\rm Sign}(k)>0$ on the other hand, one has a friction-like force. This already is a hint for the stability of the equations in the long term. The other force is like a linear force depending on the sign of the RHS of (\ref{equib}). This sign is unambiguously fixed by the requirements for stabilization. Let us remember that the coefficients of $P_{s}(X)$ start becoming positive, with decreasing negative $\tau$, starting from the highest order term, and we have shown that there is a unique positive root of $P_{s}(X)$ if there is at least one sign change in the coefficients. Thus the requirement for a unique positive root also requires that the sign of the coefficient of the highest order term be positive, this would mean that the derivative of the stabilization polynomial at the equilibirum point is positive since the positive root is the largest root and the polynomial will either increase or decrease depending on the sign of the coefficient of the largest order term. Therefor
the sign of the RHS of (\ref{equib}) is negative. The second force is an attractive linear force in the linear approximation. Hence the $C$ equations are linearly stable near the stabilization fixed point.

To try to understand what could happend in the non-linear regime let us remember again we have shown,  given the conditions on $\tau$, that there is a unique positive root $X_{0}$ of $P_{s}(X)$. This would mean that we  have $P_{s}(X)<0$ for $0<X<X_{0}$ and $P_{s}(X)>0$ for $X>X_{0}$ since the coefficient of the highest order term is positive. We will confine ourselves to   ${\rm Sign}(k)>0$. 

{\bf Case A:} Let us pick a point in the region $0<X<X_{0}$ as our initial data point. Here $\ddot{C}$ is initially negative if initially $\dot{C}>0$ ($C$ getting larger $X$ getting smaller) and hence $\dot{C}$ will become negative at some time in the future (since $\ddot{C}$ will never change sign until $\dot{C}$ becomes negative). When this transition happens we would have $C$ getting smaller and $X$ getting larger, so this initial data eventually turns toward the equilibrium point $X_{0}$ with an initial data equivalent to, $0<X<X_{0}$ and $\dot{C}<0$. Thus we should only consider this. In this case the sign of $\ddot{C}$ depends on the interplay between the terms in (\ref{equic}). There are two cases: either the evolution of the system
monotonically approaches $X_{0}$ with $\dot{C}$ approaching zero from below or passes this point with $\dot{C}<0$ and becomes a problem of Case B below.

{\bf Case B:} Let us pick a point in the region $X>X_{0}$ as our initial data point. Here $\ddot{C}$ is initially positive if initially $\dot{C}<0$ ($C$ getting smaller $X$ getting larger) and hence $\dot{C}$ will become
positive at some time in the future (since $\ddot{C}$ will never change sign until $\dot{C}$ becomes positive), when this transition happens we would get $C$ getting larger and $X$ getting smaller to this initial data eventually turns towars the equilibrium point $X_{0}$ with an initial data equivalent to
$X>X_{0}$ and $\dot{C}>0$. Thus we should only consider this.  In this case the sign of $\ddot{C}$ depends on the interplay between the terms in (\ref{equic}). There are two cases: either the evolution of the system
monotonically approaches $X_{0}$ with $\dot{C}$ approaching zero from above or passes this point with $\dot{C}>0$ and becomes a problem of Case A above.

The considerations above are enough to argue that the point $X_{0}$ is an attractive fixed point: the solutions will converge to this point in the future.

\section{Conclusions}

In this work we have assumed that branes wrap democratically over a partitioning of the extra dimensions. That is they wrap around each element and in all possibilities. Then by considering winding and momentum modes we have shown that stabilization of extra dimensions is achieved. By a further assumption that the dilaton will never enter the strong coupling regime in the future, the resulting constraint connects the coupling parameter of the dilaton to the number of
dimensions of the observed space. String theory input restricts this relation such that the upper inclusive bound on the number of dimensions of our everyday world is three.  

We have also included in this work the contribution of the ordinary presureless matter living in the observed space and have shown that this does not alter the results. Unfortunately things are different
for other modes of matter and we expect that there will be no stabilization in our scheme for radiation and
vacuum energy. However as shown in \cite{art1} it could be the case that the extra dimensions contract during radiation phase. This might compensate for the increase in the inflationary period. If this happens, as shown in this paper, the internal dimensions would stabilize after the photons decouple and ordinary matter domination starts. It is not clear however how to embed naturally the present acceleration of the universe in this scenario such that the size of internal dimensions are stabilized or change rather very slowly. Nevertheless, the democratic principle and symmetric decimation ideas we have made use of here could also render these future works amenable to reasonable detail. A work to study the effects of the early inflation and radiation era's on the model presented here is in progress.

It is rather interesting that Descartes, the one who introduced the very concept of co-ordinates has a say in this matter via a very simple mathematical rule known as "the Descartes' sign rule" to determine the number of positive roots of a polynomial.

\end{document}